\documentclass[onecolumn,showpacs,a4paper,rmp,superscriptaddress]{revtex4}
\usepackage{graphicx}
\usepackage{amsmath}
\usepackage{amssymb}
\usepackage{dcolumn}

\begin{document}

\title{High-temperature series expansions for the $q$-state Potts model on a
  hypercubic lattice and critical properties of percolation}
\date{\today}

\author{Meik Hellmund}
\email{Meik.Hellmund@math.uni-leipzig.de}
\affiliation{Mathematisches Institut, Universit{\"a}t Leipzig,
Augustusplatz 10/11, D-04109 Leipzig, Germany}
\affiliation{Centre for Theoretical Sciences (NTZ) of the Centre for
Advanced Study (ZHS), Universit{\"a}t Leipzig,\\
Emil-Fuchs-Str.\ 1, D-04105 Leipzig, Germany}
\author{Wolfhard Janke}
\email{Wolfhard.Janke@itp.uni-leipzig.de}
\affiliation{Institut f{\"u}r Theoretische Physik, Universit{\"a}t Leipzig,
Augustusplatz 10/11, D-04109 Leipzig, Germany}
\affiliation{Centre for Theoretical Sciences (NTZ) of the Centre for
Advanced Study (ZHS), Universit{\"a}t Leipzig,\\
Emil-Fuchs-Str.\ 1, D-04105 Leipzig, Germany}

\begin{abstract}
We present results for the high-temperature series expansions of the susceptibility
and free energy of the $q$-state Potts model on a $D$-dimensional hypercubic
lattice  $\mathbb{Z}^D$
for arbitrary values of $q$. The series are up to order 20 for dimension $D\leq3$, order 19
for $D\leq 5$ and  up to order 17 for arbitrary $D$.
Using the $q\rightarrow 1$ limit of these series,
 we estimate the percolation threshold $p_c$ and critical exponent
$\gamma$  for bond percolation in different dimensions.
We also  extend the  $1/D$
 expansion of  the critical coupling 
 for arbitrary values of $q$ up to order $D^{-9}$. 
\end{abstract}

\maketitle

\section{Introduction}

Since many years the $q$-state Potts model \cite{Wu-Potts} has served as  
paradigmatic system for the study of phase transitions and critical phenomena.  
Fortuin  and Kasteleyn reformulated it as a random cluster model in which
$q$ appears as a general parameter no longer restricted to $q=2,3,\dots$
and where bond percolation appears as the special case $q=1$. 

Systematic series expansions have a long history \cite{domb3} as a method to 
analyse lattice models of phase transitions. In the case of Potts models, 
high-temperature series for the susceptibility and free energy
with $q$ as a free parameter have been obtained for 
$D$-dimensional hypercubic lattices in 
\cite{moraal} (free energy only) and \cite{schreider94}. 
We extend these series considerably and use them for 
a scan of the phase diagram in $(q,D)$-space. 
A detailed analysis is presented for the case 
of bond percolation ($q=1$) on higher-dimensional hypercubic lattices,
extending the analysis of 15th order
percolation series in arbitary dimensions 
of \cite{adler90}.  

We shortly review the model and its cluster formulation in 
Sect.~\ref{sec:model} and the star-graph expansion 
technique in Sect.~\ref{series}. In the following sections we present
results for general parameter $q$ in two  (Sect.~\ref{2d}) 
and higher   (Sect.~\ref{higherd}) dimensions. 
Section~\ref{larged} treats the large-dimensionality 
expansion of the critical coupling, and in  
Sect.~\ref{perc} the case of  bond percolation, i.e., the limit $q=1$ 
is analyzed. We end with a short summary in Sect.~\ref{conc}.

\section{The $q$-state Potts model}
\label{sec:model}

Let $G=(V,E)$ be an  undirected graph, that is a set  $V$ of vertices 
together with a set $E$  of edges defining which pairs of vertices 
are considered ``nearest neighbors''.  
Put a  spin $s_i$ on each vertex $i\in V$
which can take $q$ different values.     
The $q$-state Potts model (see, e.g., \cite{Wu-Potts} for a review) 
on this graph is then defined by 
 the Hamiltonian
\begin{equation}
  \label{eq:h1}
  H = -J \sum_{(ij)\in E} \delta(s_i,s_j),
\end{equation}
where $\delta(.,.)$ is the Kronecker delta symbol.
The Fortuin-Kasteleyn
representation shows that the partition function $Z_G=\sum_{\{s_i\}} e^{-\beta H}$
for a finite graph $G$ is
actually a polynomial in $q$ and $w=e^{\beta J}-1$:
\begin{equation}
  \label{eq:z1}
  Z_G (q, w) =  \sum_{C\subseteq E} q^{N_c(C)} w^{|C|}.
\end{equation}
Here the sum is over all $2^{|E|}$ clusters, i.e. subsets $C$ of the set $E$
of edges;
$N_c(C)$ is the number of connected components where each isolated vertex
is counted as one component and $|C|$ is the number of edges of $C$.
This allows one to study the model for arbitrary values
of $q \notin \{2,3,\ldots\}$ which can no
longer be represented by a local interaction of spin degrees of 
freedom of the original Hamiltonian (\ref{eq:h1}). 
Of special interest in statistical physics are the
limits $q\to1$ and $q\to0$ which can be shown to be related to percolation
and tree percolation on the graph $G$, respectively.

A two-point correlation function can be defined in the cluster representation simply by
\begin{equation}
  \label{eq:corr}
  D(i,j) = \frac{1}{Z_G} \sum_{C_{ij}\subseteq E} q^{N_c(C)} w^{|C|},
\end{equation}
where the cluster sum is restricted to those clusters $C_{ij}$ in which
 the vertices
$i$ and $j$ belong to the same connected component.
The susceptibility related to this correlator,
\begin{equation}
  \label{eq:sus}
 \chi_G(q,w) = \frac{1}{|V|}\sum_{i,j\in V} D(i,j),
\end{equation}
equals  for $q=2,3,\ldots$ the usual magnetic susceptibility 
of the $q$-state Potts model in the high-temperature phase (i.e., as long 
as $\langle s_i\rangle=0$).  

In the case $q=1$ the weight of a cluster in the partition sum 
depends only on its number of edges. This case therefore describes 
bond percolation where the local  bond  probability 
is given by $p = w/(w+1)$. It easily follows that here
the  susceptibility $\chi(q=1,w)$ measures the average size
of the percolation cluster.

The  Potts partition function $Z_G(q,w)$  is essentially equivalent 
to the Tutte or dichromatic polynomial of the graph $G$, see,
e.g., \cite{sokal-2005} and references therein. 
It encodes a lot of combinatorical properties of the
graph such as the number of spanning trees, the number of spanning forests, 
the number of possible
acyclic orientations, or the number of vertex colourings with $q$ different 
colours using different colours on neighbouring vertices. The last
mentioned number, for example,  equals
  $\lim_{\beta J\rightarrow -\infty} Z_G(q,e^{\beta J}-1) = Z_G(q,-1)$, 
which is the number of ground states of the antiferromagnetic $q$-state 
Potts model on the graph $G$.
 
In statistical physics, on the other hand, the interest in these models is
related to the existence of a phase transition between 
a disordered high-temperature (low-density in case of percolation clusters)
and an ordered
low-temperature
(large $w$)  phase
 if we go to infinite graphs
like the $D$-dimensional hypercubic  lattice $G=\mathbb Z^D$. 
Depending on $q$ and $D$, these transitions can be first or second order.
Whereas in $D=2$ dimensions many exact results are available showing that
the second-order nature changes to first-order for $q > 4$, in $D=3$ and
higher dimensions there is strong numerical evidence that the transition is of
first order for all integer values $q > 2$.
In the case of a second-order phase
transition at $\beta_c$, the behavior of the susceptibility \eqref{eq:sus} is
characterized by the critical exponent 
$\gamma$: $\chi_G \sim |\beta-\beta_c|^{-\gamma}$.

\section{High-temperature series expansions} 
\label{series}
Series expansions are an important tool for the extraction of information on
statistical systems which is exact (up to the order of the expansion) in the
thermodynamic limit. There exist different well-established 
methods \cite{domb3} for the systematic
generation of high-temperature series expansions, as for example
the linked cluster and the star-graph method. The latter one got its name 
from the fact that  only
1-vertex-irreducible graphs contribute to the series. 
In an attempt to study systems with quenched disorder via series expansions
\cite{Hellmund:2005b,hellmund05a,Hellmund:2003,Hellmund:2002,Hellmund:2002a}
we used this method  and 
developed a  comprehensive software toolbox for generating and enumerating
star graphs which is re-used in this work.  

The basic idea of the star-graph method 
is to assemble the  value of some extensive thermodynamic
quantity $F$ on a large or even infinite graph from its values on subgraphs:
Graphs constitute a partially ordered set under the ``subgraph'' relation.
Therefore, for every function  $F(G)$ defined on the set of graphs
exists another function $W_F(G)$ such that for all graphs $G$
\begin{equation}
  \label{eq:2}
   F(G) = \sum_{g \subseteq  G} W_F(g),
\end{equation}
and this function can be calculated recursively via
\begin{equation}
  \label{eq:3}
  W_F(G) = F(G) -  \sum_{g \subset   G} W_F(g).
\end{equation}
This gives for an infinite (e.g. hypercubic) lattice
\begin{equation}
  \label{eq:4}
  F(\mathbb{Z}^D) = \sum_G (G:\mathbb{Z}^D)\, W_F(G),
\end{equation}
where $(G:\mathbb{Z}^D)$ denotes the weak
embedding number of the graph $G$ in the given
lattice structure~\cite{martin74}.

The following observation makes this a useful method:
Let $G$ be  a  graph with an articulation vertex
where two star subgraphs $G_{1,2}$ are glued together.
Then $W_F(G)$ vanishes if
\begin{equation}
  \label{eq:sg}
F(G) = F(G_1) + F(G_2).
\end{equation}
An observable $F$ for which eq.~(\ref{eq:sg}) is true on arbitrary graphs
with articulation points allows a star-graph expansion. Then all non-star graphs
have zero weight $W_F$ in the sum eq.~(\ref{eq:4}).
It is easy to see that the (properly normalized) free energy $\ln Z$
has this property and it can be shown  that the inverse
susceptibility $1/\chi$ has it, too.
 
The weak embedding numbers  $(G:\mathbb Z^D)$ of a graph $G$ into the
hypercubic lattice are counted using a refined version 
of the backtracing algorithm by
Martin \cite{martin74}, making use of a couple of simplifications for
bipartite hypercubic lattices ${\mathbb Z}^D$.
Figure~\ref{fig:emb} shows two typical results. A star graph of order $n$ can use
at most $\lfloor\frac{n}{2}\rfloor$ dimensions of the lattice and every
embedding using $m$ dimensions appears in ${\mathbb Z^D}$   with multiplicity 
$\binom{D}{m}$.

\begin{figure}[H!t]
\unitlength6mm
\begin{minipage}{0.49\textwidth}\vspace*{0.5cm}
\begin{center}
  \begin{picture}(8,1)
\thicklines
\put(0,0){\line(1,0){8}}
\put(0,0){\line(0,1){1}}
\put(0,1){\line(1,0){8}}
\put(8,0){\line(0,1){1}}
\put(3,0){\line(0,1){1}}
\put(0,0){\circle*{.24}}
\put(1,0){\circle*{.24}}
\put(2,0){\circle*{.24}}
\put(3,0){\circle*{.24}}
\put(4,0){\circle*{.24}}
\put(5,0){\circle*{.24}}
\put(6,0){\circle*{.24}}
\put(7,0){\circle*{.24}}
\put(8,0){\circle*{.24}}
\put(0,1){\circle*{.24}}
\put(1,1){\circle*{.24}}
\put(2,1){\circle*{.24}}
\put(3,1){\circle*{.24}}
\put(4,1){\circle*{.24}}
\put(5,1){\circle*{.24}}
\put(6,1){\circle*{.24}}
\put(7,1){\circle*{.24}}
\put(8,1){\circle*{.24}}
\end{picture}
\end{center}
\small
\begin{eqnarray*}
   & 7620 \binom{D}{2} +  76851600    \binom{D}{3}+ 14650620864 \binom{D}{4}\\
 &+\; 404500471680\binom{D}{5}+ 3355519311360   \binom{D}{6}
\end{eqnarray*}
\end{minipage}
\begin{minipage}{0.49\textwidth}\vspace*{-0.5cm}
\begin{center}
  \begin{picture}(6,3)
\thicklines
\put(0,0){\line(1,0){3.9}}
\put(1,1){\line(1,0){3.9}}
\put(0,0){\line(1,1){1}}
\put(3.9,0){\line(1,1){1}}
\put(4.9,1){\line(0,1){1.3}}
\put(2.6,0){\line(0,1){1.3}}
\put(1.3,1.3){\line(1,0){2.6}}
\put(2.3,2.3){\line(1,0){2.6}}
\put(1.3,1.3){\line(1,1){1}}
\put(2.6,1.3){\line(1,1){1}}
\put(3.9,1.3){\line(1,1){1}}
\put(0,0){\circle*{.24}}
\put(1,1){\circle*{.24}}
\put(1.3,0){\circle*{.24}}
\put(2.6,0){\circle*{.24}}
\put(3.9,0){\circle*{.24}}
\put(2.3,1){\circle*{.24}}
\put(3.6,1){\circle*{.24}}
\put(4.9,1){\circle*{.24}}
\put(1.3,1.3){\circle*{.24}}
\put(2.6,1.3){\circle*{.24}}
\put(3.9,1.3){\circle*{.24}}
\put(2.3,2.3){\circle*{.24}}
\put(3.6,2.3){\circle*{.24}}
\put(4.9,2.3){\circle*{.24}}
\end{picture}
\end{center}
\small
\begin{eqnarray*}
   &12048 \binom{D}{3}+  396672\binom{D}{4} +  2127360\binom{D}{5}+
2488320\binom{D}{6}
\end{eqnarray*}
\end{minipage}

  \caption{Two star graphs of order 19 (left) and 17 (right)
and their weak embedding numbers. Due to its topology, the result is complete
(valid for arbitrary dimension $D$) for the graph of order 17. For the graph 
of order 19 it is valid only up to $D=6$.}
  \label{fig:emb}
\end{figure}

For the symbolic calculations of the partition sum and susceptibility 
on every star graph up to a given order (ca. $80\,000$ up to order 20) 
as polynomials in $q$ and $w$,  
we developed a C\raise2pt\hbox{\tiny++}
template library using an expanded degree-sparse
representation of polynomials and series in many variables.
 The open source library GMP is used  for the
arbitrary-precision arithmetics.
Our series for the free energy $\ln Z$ and the susceptibility $\chi$
with $q$ as a free parameter 
are up to order 20 for dimension $D\leq3$, order 19
for $D\leq 5$ and  up to order 17 for arbitrary $D$. 

In order to check our algorithms we did a careful comparison with 
older published series data. Our series agree with the 10th order
susceptibility  
and free energy series for arbitrary $q$ and $D$ of
\cite{schreider94} (modulo some misprints in
their expression for $\ln Z$) as well as with the 16th order $D=2$ free energy 
series of \cite{moraal}  but disagree in several places with the 
11th order, arbitrary dimension free energy series in the last mentioned paper.
For the special case of bond percolation $q=1$ they agree with the 15th order 
arbitrary-dimension series of \cite{adler90}. In the special case of the 
three-dimensional Ising model ($q=2$, $D=3$) our 20th order series also 
conform with the 25th order series of \cite{butera02} generated by
linked-cluster expansion. They also agree with the 21th order $q=3, D=3$
partition function series of \cite{guttmann94} obtained by finite lattice
methods. In two dimensions, much longer series exist for special models. Our
20th order susceptibility series agree  with the $q=3$ and $4, D=2$ series of 
\cite{guttmann03}.

We close this section with some remarks about the analysis of the thus
generated expansions.
The estimate of critical parameters from a high-temperature series 
involves extrapolation from a finite number of exactly known coefficients 
to the asymptotic form of the function. Many such extrapolation techniques
have been developed and tested for different series and are comprehensively 
reviewed in \cite{guttmann89}. 
These extrapolation techniques are not rigorous. They make some assumptions
about the expected form of the singularity at the critical temperature.    
Usually, error estimates rely on the scatter of the results of extrapolations
with different parameters (like $[N/M]$ Pad\'e approximants  for different 
values of $N$ and $M$). This may seriously underestimate systematic errors
coming from wrong assumptions about the structure of the singularity. 

In order to get a reliable picture, we will take into account several
criteria, such as 
\begin{itemize}
\item convergence of the analysis,
\item scatter of different approximants, 
\item number of defective approximants,
\item agreement between different extrapolation methods.  
\end{itemize}

The basic methods we use are DLog-Pad\'e approximation and inhomogeneous
differential approximants (IDA) \cite{guttmann89}. 
In order to analyze confluent nonanalytic
and logarithmic 
corrections,  these methods are applied to suitably
transformed forms of the series. The parameters of these transformations are 
fine-tuned according to the criteria listed above, a technique 
pioneered in \cite{adler90}.

\section{Test: Two dimensions}
\label{2d}
The Potts model in two dimensions is exhaustively reviewed in \cite{sokal04}.
We repeat here the principal facts:
Baxter  has determined the exact free energy on ${\mathbb Z}^2$ along the two
curves
\begin{eqnarray}
  \label{eq:fl}
  w_F &=& \sqrt{q}\\ \label{eq:fa}
  w_A &=& -2 + \sqrt{4-q}
\end{eqnarray}
in  $(q,w)$ parameter space.
The curve \eqref{eq:fl} gives the locus of the ferromagnetic phase transition
and is perfectly understood in terms of conformal field
theory. Critical exponents are exactly known, e.g.
\begin{equation}
  \label{eq:gam2}
\gamma = \frac{4+3k^2}{12k-6k^2} \text{\ \ \ where\ \ \ } \sqrt{q} = -2 \cos(\pi/k).
\end{equation}

The curve \eqref{eq:fa} is conjectured to be the locus of the
antiferromagnetic transition which agrees with the known facts for  $q=2,3,4$.

\begin{figure}[H!t]
  \begin{center}
\includegraphics[scale=.6]{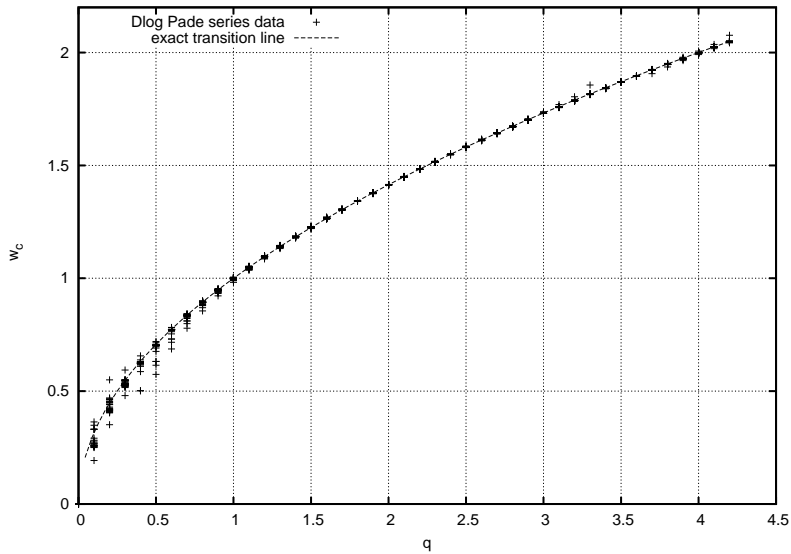}
    \caption{Critical coupling $w_c$ as function of $q$ in 2
      dimensions.}
    \label{fig:1}
  \end{center}
\end{figure}

\begin{figure}[H!t]
  \begin{center}
\includegraphics[scale=.4,angle=-90]{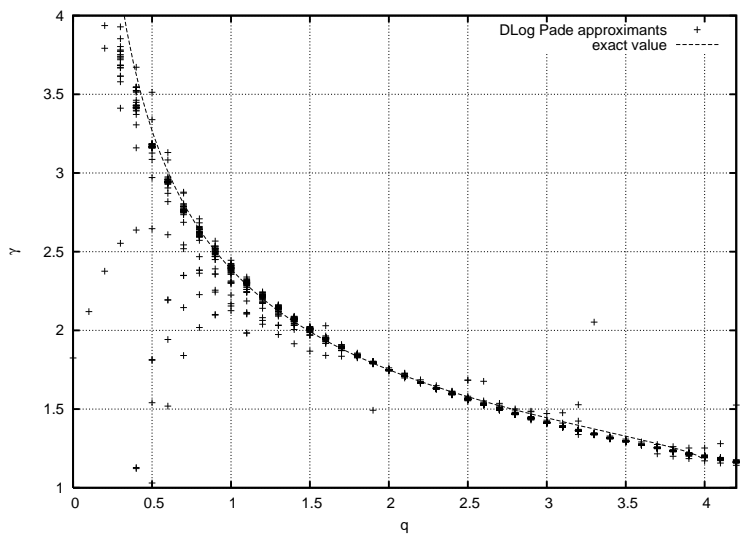}
    \caption{Critical exponent $\gamma$ as function of $q$ in 2
      dimensions.}
    \label{fig:2}
  \end{center}
\end{figure}

An analysis of the susceptibility series of order 20 using  Pad\'e
approximants to the logarithmic derivative of the susceptibility
(DLog-Pad\'e approximants) 
is shown for the ferromagnetic case in Figs.~\ref{fig:1} and \ref{fig:2} 
where around 15--25 different
$[N,M]$  approximants are plotted for each value of $q$. The continuous
lines present the exact results \eqref{eq:fl} and \eqref{eq:gam2}. 
We find a  quite satisfactory agreement also for noninteger values of $q$,
however, the scatter of different approximants and the number of defective
approximants increases for smaller values of $q$. This may be related to the
fact that  the $q\rightarrow 0$ limit is a quite intricate 
multicritical point \cite{sokal04}.

\section{Higher dimensions: Overview}
\label{higherd}
\begin{figure}[H!t]
  \begin{center}
\includegraphics[scale=.4,angle=-90]{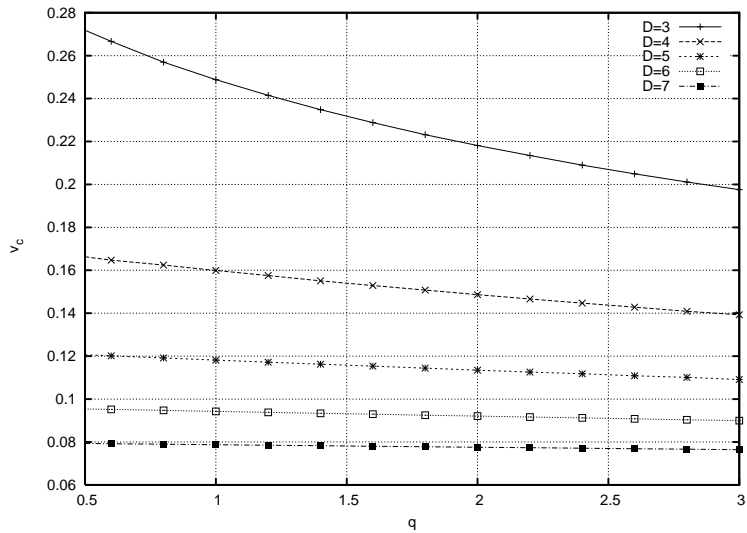}
    \caption{Critical coupling $v_c$ as function of $q$ in different
      dimensions.}
    \label{fig:5}
  \end{center}
\end{figure}

\begin{figure}[H!t]
  \begin{center}
\includegraphics[scale=.4,angle=-90]{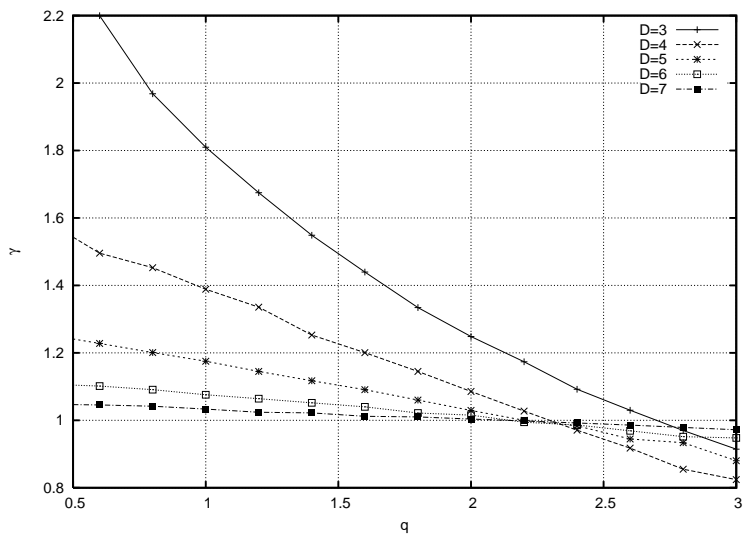}
    \caption{Critical exponent $\gamma$ as function of $q$ in different
      dimensions.}
    \label{fig:6}
  \end{center}
\end{figure}

The explicit $q$- and $D$-dependence of our series allows us to get a fast
overview over the phase structure in a large parameter range. Figures
\ref{fig:5} and \ref{fig:6} show the results of a  DLog-Pad\'e analysis of
the susceptibility series for the critical coupling  
$v = (e^{\beta J}-1)/(e^{\beta J} -1+q) = w/(w+q)$ and exponent $\gamma$.

Several
remarks are in order:
\begin{itemize}
\item At the upper critical dimension ($D=4$ for $q=2$, $D=6$ for $q=1$) the
 DLog-Pad\'e analysis finds a value $\gamma\approx1.08$ slightly larger than $1$. This
 effect is well-known and due to the existence of logarithmic corrections.
\item For $D\geq 3, q\geq3$ the phase transition is of first order and the
  correlation length remains finite. In analysing high-temperature series by 
  Pad\'e approximants or similar techniques, the
  approximant will provide an analytic continuation beyond the first-order
  transition temperature $T_0$ into a metastable region on a pseudo-spinodal line 
  with a singularity $T^*_c < T_0$ and effective exponents at $T^*_c$. Our $v_c$ in
  this region is therefore an upper bound of the real transition point.
\item For $D=3$, the curve $\gamma(q)$ passes $\gamma=1$ at $q\approx2.7$,
  marking the onset of a first-order transition. This is consistent with slightly
  smaller estimates in the literature coming from Monte Carlo computer
  simulations and field theoretic calculations (see \cite{grollau01} and
  references therein). 
  It would be interesting to see if the crossover point can be understood as a kind
  of tricritical point in the $(q,v)$ parameter space.
\end{itemize}

Since high-temperature series alone are not sufficient to analyse first-order
transitions (and much longer series exist for the special case
of the three-dimensional Ising model $q=2, D=3$), we concentrate our analysis in
the second part of the paper on the $q=1$ case of bond percolation.

\section{Large-Dimensionality Expansion}
\label{larged}
The dimension of the lattice enters the star-graph series expansion 
only through the embedding numbers (eq.~\eqref{eq:4}) and we keep 
it as a free parameter in our series (see Fig.~\ref{fig:emb}). 
This allows one~\cite{fishersingh}  to extract a large-$D$ expansion for the
critical coupling from the susceptibility series
by iteratively solving the critical point equation
$1/\chi(D,v_c) = 0$ in terms of  $v_c$ as series in $1/D$. 
This gives the large-$D$ expansion of $v_c$ up
to order $D^{-n}$ if $\chi$ is known as series in $v$ up to order $2n$.
For the $O(N)$ model the first terms 
of the $1/D$ expansion of the critical coupling for arbitrary $N$ were
obtained by Gerber and Fisher \cite{gerberfisher}, who also demonstrated 
by solving the large $N$ limit that this expansion is only asymptotic.  

As far as we know, our analogous series for the $q$-state Potts model with
arbitrary $q$ is new. For the percolation ($q=1$) case the first 5 orders
were calculated in \cite{gauntruskin}. 
We give the series for $v_c$ in terms of the branching number $\sigma$
which is one less than the coordination number of the lattice and therefore
for the hypercubic lattice
$\sigma = 2D-1$,

\begin{eqnarray}
  \label{eq:inv}\nonumber
v_c(q,\sigma) = && \frac{1}{\sigma} \left[
1+\frac{8-3 q}{2   \sigma ^2} + \frac{3 (8-3 q)}{2 \sigma ^3}
+\frac{3 \left(68-31 q+q^2\right)}{2 \sigma^4}
+\frac{8664-3798 q-11 q^2}{12 \sigma^5} \right. \\ \nonumber&& \quad\quad
+\frac{78768-36714 q+405 q^2-50 q^3}{12\sigma ^6}
+\frac{1476192-685680 q-2760 q^2-551 q^3}{24\sigma ^7}
\\  && \quad\quad \left.
+\frac{7446864-3524352 q-11204 q^2-6588 q^3-9 q^4}{12 \sigma ^8}
+ \cdots \right].
\end{eqnarray}

\begin{figure}[H!bt]
  \begin{center}
\includegraphics[scale=.4,angle=-90]{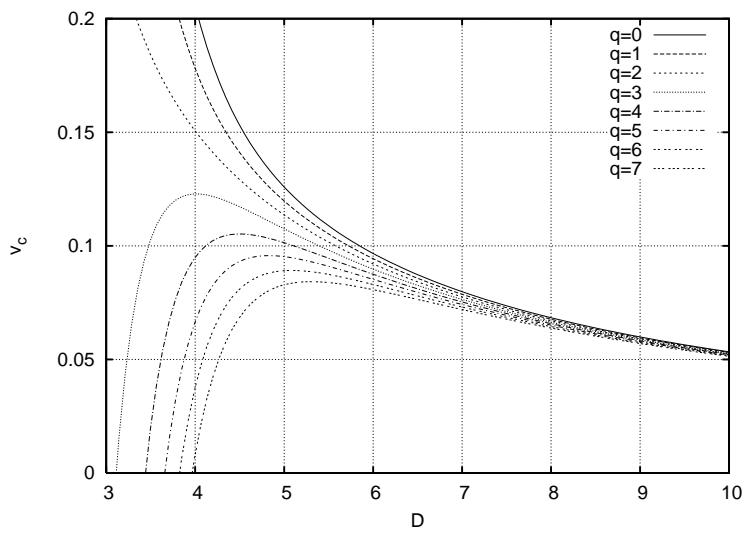}
    \caption{ Plots of eq.~(\ref{eq:inv}) (using all available terms, cf.\ text) 
    for different values of $q$.}
    \label{fig:8}
  \end{center}
\end{figure}

By using $\sigma$ as expansion variable we see that the first correction to
the Bethe lattice result $v_c=1/ \sigma$ is absent for all values of $q$.
Figure~\ref{fig:8} shows plots of the critical coupling as obtained from 
eq.~(\ref{eq:inv}) by simply summing up all terms. Since these expansions
are asymptotic series, for relatively small $D$ a more careful treatment with
appropriately adapted maximal order would actually be necessary, as discussed
below for the percolation case $q=1$. Of course, in the case of first-order
phase transitions, the expansion (\ref{eq:inv}) yields the pseudo-spinodal 
transition point in the meta-stable region.

\section{$q\to1:$ Percolation}
\label{perc}

The $q\to1$ limit of the Potts model describes bond percolation with 
local bond probability $p = w/(w+1) = v$.
Standard scaling theory \cite{staufferbook} describes the number of clusters
of size $s$ for large $s$ near the percolation threshold $p_c$ by
\begin{equation}
  \label{eq:scalperc}
  n(s,p) \sim s^{-\tau}  f( (p_c-p) s^\sigma),
\end{equation}
 introducing the critical exponents $\tau$ and $\sigma$.
The $q\to1$ limit of the Potts model susceptibility describes the average
cluster size and $\frac{\partial \ln Z}{\partial q}$ the number of clusters.
  The respective  critical exponents are related to the
scaling law eq.~(\ref{eq:scalperc})  by
\begin{eqnarray}
  \label{eq:susperc}
  \gamma&=&\frac{3-\tau}{\sigma},\\
 2-\alpha &=& \frac{\tau-1}{\sigma}.
\end{eqnarray}

A throughout anaysis of percolation series in higher dimensions was done
some years ago by 
\cite{adler90}.
Since we extended the series from 15 to 17 (for $D\ge6$), 19 (for $D=4,5$) 
and  20 (in $D=3$)  terms,  a careful
re-analysis seems to be in place 
(for extensive percolation series expansions in two dimensions, see
\cite{conw-guttm} and, recently, \cite{JensenZiff}).

\subsection{Up to the upper critical dimension}

\begin{table}[H!t]
  \centering
\caption{Percolation thresholds (bond percolation on $\mathbb Z^D$)
 and critical exponents.}
\begin{ruledtabular}
  \begin{tabular}{c|r|d|d|d|d}
$D$ & Methods & 
\multicolumn{1}{c|}{$p_c$} & 
\multicolumn{1}{c|}{$\sigma$} & 
\multicolumn{1}{c|}{$\tau$} & 
\multicolumn{1}{c}{$\gamma=\frac{3-\tau}{\sigma}$}\\ \hline
2 & exact, CFT & \multicolumn{1}{c|}{$\frac{1}{2}$} &
\multicolumn{1}{c|}{$\frac{36}{91}$} & 
\multicolumn{1}{c|}{$\frac{187}{91}$} &
\multicolumn{1}{c}{$\frac{43}{18}$}\\
 \hline
 & MC \cite{ZiffStell}& 0.248\,812(2)& 0.453(1)&2.189(1)&1.795(5)\\
& HTS \cite{adler90} &0.248\,8(2) &&& 1.805(20)\\
3 & MC \cite{lorenz98}& 0.248\,812\,6(5)&&&\\ 
 & MCS \cite{ballesteros-1999}&&0.4522(8) & 2.18906(6)& 1.7933(85) \\
& MCS \cite{dengBloete}&&0.4539(3)& 2.18925(5)& 1.7862(30)\\
 & {\bf present work} & 0.248\,91(10) &&& 1.804(5)  \\
\hline
  & HTS \cite{adler90}&0.160\,05(15)& &&1.435(15)\\
 & MCS \cite{ballesteros-1997}&&&&1.44(2)\\
4 & MC \cite{paulziff-2001}&0.160\,130(3)& & 2.313(3)&\\
  & MC \cite{grassberger-2002} & 0.160\,131\,4(13)& &&\\
  &   {\bf present work} & 0.160\,08(10) &&& 1.435(5) \\
\hline
 &HTS \cite{adler90}& 0.118\,19(4)&&&1.185(5)\\
5 & MC \cite{paulziff-2001}& 0.118\,174(4)&&2.412(4)&\\
 & MC \cite{grassberger-2002}& 0.118\,172(1)&&&\\
 & {\bf present work} & 0.118\,170(5) &&& 1.178(2) \\
\hline
 & RG \cite{essamgaunt} &
 & \multicolumn{1}{c|}{$\frac{1}{2}$} &
\multicolumn{1}{c|}{$\frac{5}{2}$} & 
\multicolumn{1}{c}{$\chi \sim t^{-1} |\ln t|^{\delta}, \; \delta=\frac{2}{7}$}\\
 6 & HTS \cite{adler90}&0.094\,20(10) &&&\\
 & MC \cite{grassberger-2002} & 0.094\,201\,9(6) & && \\
 & {\bf present work} & 0.094\,202\,0(10) & && \delta=0.40(2) \\
\hline
$>6$ &  RG& & \multicolumn{1}{c|}{$\frac{1}{2}$} &\multicolumn{1}{c|}{$\frac{5}{2}$} & 1\\
  \end{tabular}
\end{ruledtabular}\\[3mm]
  \label{tab:p1}
MC = Monte Carlo, MCS = Monte Carlo, {\bf site}
percolation, RG = Renormalization group, HTS = High-temperature series
\end{table}

The divergence of the susceptibility (mean finite cluster size)
is expected to be of the form
\begin{equation}
  \label{eq:sp}
  \chi (p) = A (p_c-p)^{-\gamma} [1 + a_1 (p_c-p)^{\Delta_1} + \ldots ]
\end{equation}
for $D\neq6$ and
\begin{equation}
  \label{eq:sp1}
  \chi (p) = A (p_c-p)^{-\gamma} |\ln(p_c-p)|^\delta + \ldots
\end{equation}
 at the upper critical dimension $D = 6$ with $\gamma=1$ and $\delta = 2/7$.
For $D>6$, one expects a  Gaussian fixed point with  $\gamma=1$. 

We analyse the series using DLog-Pad\'e approximants,
inhomogeneous differential approximants and the methods termed M1 and M2 
from \cite{adler91,adler90} especially  
tailored to take confluent nonanalytic correction terms into account.
 The method M1 uses DLog-Pad\'e approximants to 
\begin{equation}
  \label{eq:m1}
  F(p) = (p_c-p) \frac{d\chi}{dp} -\gamma \chi(p),
\end{equation}
    which has a pole at $p_c$ with residue $\Delta_1 -\gamma$. For a given trial value
    of $p_c$ the graphs of $\Delta_1$ versus input $\gamma$ are plotted for different
    Pad\'e approximants and by adjusting $p_c$ a point of optimal convergence is
    searched.  

The M2 method starts with a transformation of the series in $p$ into a series
in $y=1-(1-p/p_c)^{\Delta_1}$, and then Pad\'e approximants to
\begin{equation}
  \label{eq:M2}
  G(y) = \Delta_1 (1-y) \frac{d}{dy} \ln F(y)
\end{equation}
are calculated which should converge to $\gamma$ as $y\to1$, i.e. $p \to p_c$.
These methods are especially useful when taken as biased
    approximants with a given value of $\gamma$ or $\Delta_1$ as input. 
Our results are compiled in Table~\ref{tab:p1}, where we quote for comparison
also previous estimates based on Monte Carlo simulations and analyses of 
shorter series expansions.

Generally the series analysis appears to be more difficult for 
lower dimensions. 
The $p_c$ we quote for $D=3$ is an average over a large number (ca.\ 200) of
Pad\'e and IDA approximants. The general pattern in the scatter of data 
is that larger values of $p_c$ corresponds with larger values of the
critical exponent $\gamma$. The value for  $\gamma$ we quote as our result in three
dimensions is obtained by
biasing the M1 and M2 analyses with the more precise $p_c$ value 
from Monte Carlo simulations. 
Moving on to $D=4$ and $D=5$ dimensions, nothing spectacular happens and
in particular we do not observe any indication of logarithmic corrections 
to the critical behavior in $D=5$, in accordance with 
renormalization group expectations and also with newer simulations,
cf.\ the careful discussion of five-dimensional percolation in \cite{fortunato04}. 

Due to the theoretically expected multiplicative logarithmic 
corrections \eqref{eq:sp1} at the upper critical dimension 
a special treatment of the case $D=6$  is needed. The analysis of logarithmic 
corrections of the general form
\begin{equation}
  \label{eq:log}
  f(p) \sim (p_c-p)^{-\gamma} |\ln (p_c-p)|^{\delta}
\end{equation}
is possible by \cite{adler81} calculating approximants for 

\begin{equation}
  \label{eq:log2}
F(p) = (p_c-p) \ln(p_c-p) \left[\frac{f'(p)}{f(p)}-\frac{\gamma}{p_c-p}\right],
\end{equation}
where one expects for singularities of the form (\ref{eq:log}) that 
$\lim_{p\to p_c} F(p) = \delta$. Using this method biased to $\gamma=1$ we 
find the values for $p_c$ and $\delta$ listed in Table~\ref{tab:p1}.

\subsection{The percolation threshold in large dimensions}
\label{sec:percdim}

In \cite{grassberger-2002}, Grassberger compared his Monte Carlo results for $p_c$ with
 a large-$D$ expansion up to order 5 in $1/D$ \cite{gauntruskin}. Since we extended
this expansion up to order 9,
a new comparison may be in order. This expansion is believed 
to be only asymptotic. Therefore we stop summing up terms 
of the expansion if the next term is larger than the current one.
This recipe results in using only the terms up to order 7 in $D=5$ and
using all available terms for $D>5$.
We also analyzed our  susceptibility series at fixed $D$ 
using differential approximants and
the M1 and M2 methods.  Above the critical dimension, the M1 and M2
methods are used in a variant biased to $\gamma=1$. Our results are collected
in Table~\ref{tab:3}.

\begin{table}[H!t]
\caption{\label{tab:3} Bond percolation thresholds for hypercubic lattices 
$\mathbb Z^D$ for dimensions $D\geq5$.}
\begin{ruledtabular}
  \begin{tabular}{r|l|l|l}
  $D$ & Present HT series & MC \cite{grassberger-2002} &
Present $1/D$-expansion eq.~(\ref{eq:inv})\\
\hline
  5  & 0.118\,170(5)      & 0.118\,172(1)      &   0.118\,149   \\
  6  & 0.094\,202\,0(10)  & 0.094\,201\,9(6)   &   0.094\,354\,3  \\
  7  & 0.078\,682(2)      & 0.078\,675\,2(3)   &   0.078\,688\,1 \\
  8  & 0.067\,712(1)      & 0.067\,708\,39(7)  &   0.067\,708\,0  \\
  9  & 0.059\,497(1)      & 0.059\,496\,01(5)  &   0.059\,495\,1 \\
 10  & 0.053\,093\,5(5)   & 0.053\,092\,58(4)  &   0.053\,092\,13 \\
 11  & 0.047\,950\,3(1)   & 0.047\,949\,69(1)  &   0.047\,949\,47\\
 12  & 0.043\,724\,1(1)   & 0.043\,723\,86(1)  &   0.043\,723\,76\\
 13  & 0.040\,187\,7(1)   & 0.040\,187\,62(1)  &   0.040\,187\,57\\
 14  & 0.037\,183\,8(1)   &                    &   0.037\,183\,68\\
  \end{tabular}
\end{ruledtabular}
\end{table}

For higher dimensions our large-dimensionality expansion is in perfect 
agreement with Grassberger's recent Monte Carlo (MC) data \cite{grassberger-2002},
whereas direct analyses of the susceptibility series in the respective
dimension give slightly larger estimates for $p_c$. 
An obvious outlier is $D=6$ where the cut-off criterion for the $1/D$-expansion
apparently does not work (taking artificially one term less would significantly 
improve the estimate).

\subsection{Mean cluster number and moments of
the cluster number}

Our series for the free energy $\ln Z$ do not allow a reliable estimation of
the critical exponent $\alpha$. This phenomenon is already known from, e.g., Ising
model series. One needs higher derivatives $\frac{\partial^n\ln Z}{\partial p^n}$ in  order
to see a singular contribution which  shortens the series and
due to the nature of the  non-singular background the usual extrapolation
methods give much worse results than, e.g., for the susceptibility series.

On the other hand, we can determine some interesting non-universal
quantities such as the mean number of clusters per site,
\begin{equation}
  \label{eq:1}
   \langle n_c\rangle= \lim_{||G|| \rightarrow \infty} \langle N_c \rangle/||G||,
\end{equation}
and
its $n$th moments. Here, $||G||$ is the number of sites of the graph $G$. 
Using $w=q\frac{p}{1-p}$ and denoting the number of edges of the graph by $|G|$,
we rewrite the partition sum~(\ref{eq:z1}) as 
\begin{equation}
  \label{eq:zz1}
  Z_G(q,p) = \frac{1}{(1-p)^{|G|}}\sum_C  q^{N_c(C)+|C|}\; p^{|C|}\; 
(1-p)^{|G|-|C|},
\end{equation} 
implying that 
\begin{equation}
  \label{eq:5}
  \langle (N_c + |C|)^n\rangle = \lim_{q\to1}  \frac{1}{Z}
  \left(q\frac{\partial}{\partial  q}\right)^n Z.  
\end{equation}

The mean number of active bonds
per side $n_b = |C|/||G||$ is for a $D$-dimensional hypercubic lattice 
given by $\langle n_b  \rangle = pD$ and we get
\begin{equation}
  \label{eq:6}
  \langle n_c \rangle  =  \lim_{q\to 1} \frac{\partial f}{\partial  q} -pD
\text{\ \ \ \ and\ \ \ \ }  ||G||\left[ 
\langle (n_c+ n_b)^2\rangle - \langle n_c+n_b\rangle^2 \right]=
\lim_{q\to 1} \frac{\partial}{\partial q} \left(q 
\frac{\partial f}{\partial  q}\right),
\end{equation}
where $f$ is the free energy $\ln Z $ per site. 

By using Pad\'e extrapolation to estimate $\partial f/\partial q$ and
$\partial^2 f/\partial q^2$ at the percolation threshold $p_c$, we obtain the
estimates for the mean cluster numbers and the variances eq.~\ref{eq:6} 
given in 
Table~\ref{tab:4}. 
As usual, the error estimates in Table~\ref{tab:4} characterize the scatter 
of Pad\'e approximants and 
do not include systematic errors which are presumably much larger.

\begin{table}[H!t]
\caption{\label{tab:4} Mean cluster number and fluctuation.}
\begin{ruledtabular}
  \begin{tabular}{c|l|l}
Dim. & $\langle n_c \rangle$& 
$||G||\left[\langle (n_c+ n_b)^2\rangle - \langle n_c+n_b\rangle^2\right]$\\ 
\hline
2 & 0.097\,9(1)   &  0.161(3)        \\
3 & 0.272\,89(3)  &  0.029\,7(5)     \\
4 & 0.365\,494(2) &  0.007\,56(2)    \\
5 & 0.411\,852(5) &  0.003\,04(3)    \\
6 & 0.436\,327(5) &  0.001\,64(1)     \\
\end{tabular}
\end{ruledtabular}
\end{table}

For the first number in Table~\ref{tab:4}, the number of clusters per 
site $\langle n_c \rangle$ in critical $D=2$ bond percolation, 
Temperley and Lieb 
derived an exact
expression in 1971. This expression was simplified in \cite{ziff-1997} to
$\langle n_c \rangle = \frac{3\sqrt{3}-5}{2} \approx 0.098076$ which affirms
our analysis.
  For bond percolation in three dimensions, Monte Carlo simulations 
\cite{lorenzZiff98}
obtained $\langle n_c \rangle = 0.272\,931\,0(5)$. 
The fluctuations were studied in a recent Monte Carlo simulation
\cite{dengYang}. Their results 
given for $||G|| \left[ \langle n_c^2\rangle -\langle n_c\rangle^2\right]$ 
and $||G|| \left[ \langle n_c n_b\rangle -\langle n_c\rangle 
\langle n_b \rangle \right]$ together with $||G|| 
\left[ \langle n_b^2\rangle -\langle n_b\rangle^2\right] = \frac{z}{2}
p(1-p) $ where $z=2D$ is the coordination number of the lattice and $p$ 
equals $p_c$   
imply 
 $||G|| \left[\langle (n_c+ n_b)^2\rangle - \langle n_c+n_b\rangle^2\right] = 
0.164\,45(8) $ for two dimensions and 
$ 0.030\,71(45)$ for three dimensions, 
slightly larger but still consistent with our data.

\section{Summary}
\label{conc}
We successfully applied the method of high-temperature series expansion
to $q$-state Potts model on hypercubic lattices.
Modern computer facilities enable the calculation of such series while  
keeping some parameters (like $q$ or even $D$) symbolic. This allows 
one to scan a whole parameter range.
Even for the special case $q=1$ we extended the known series by several
terms and the results of the singularity analysis    
are comparable to and in good agreement with Monte Carlo data. 
By reverting the inverse susceptibility series, we derived the large-$D$
expansion for the (pseudo-) transition point of general $q$-state Potts
models up to order $D^{-9}$.
Further applications such as the study of the tree percolation limit 
$q\to 0$ are  conceivable, too.

\begin{acknowledgments}
We thank Joan Adler for discussions and help with the series analysis
and Robert Ziff for a discussion on cluster cummulants.
Support by  DFG grant No.~JA 483/17-3 and partial support from the
German-Israel-Foundation under
grant No.~I-653-181.14/1999 is gratefully acknowledged.
\end{acknowledgments}

%

\end{document}